# First Light results from PARAS: The PRL Echelle Spectrograph

Abhijit Chakraborty[1a], Suvrath Mahadevan[b,d], Arpita Roy[b,d], Fazalahmed M. Pathan[a], Vishal Shah[a], Eric H. Richardson[c], Girish Ubale[a] Rajesh Shah[a]

[a]Astronomy & Astrophysics Division, Physical Research Laboratory, Ahmedabad 380009, India; [b]Dept. of Astronomy & Astrophysics, Pennsylvania State University, University Park, PA 16802, USA; [c] University of Victoria, Victoria B.C., Canada V8W 3P6; [d]Center for Exoplanets & Habitable Worlds, Pennsylvania State University, University Park, PA 16802, USA

**ABSTRACT**

We present the first light commissioning results from the Physical Research Laboratory (PRL) optical fiber-fed high resolution cross-dispersed Echelle Spectrograph. It is capable of a single- shot spectral coverage of 3700A to 8600A at R ~ 63,000 and is under very stable conditions of temperature (0.04°C at 23°C). In the very near future pressure control will also be achieved by enclosing the entire spectrograph in a low-pressure vacuum chamber (~0.01mbar). It is attached to a 1.2m telescope using two 50micron core optical fibers (one for the star and another for simultaneous Th-Ar spectral calibration). The 1.2m telescope is located at Mt. Abu, India, and we are guaranteed about 80 to 100 nights a year for observations with the spectrograph. The instrument will be ultimately used for radial-velocity searches of exoplanets around 1000 dwarf stars, brighter than 10th magnitude, for the next 5 years with a precision of 3 to 5m/s using the simultaneous Th-Ar spectral lamp reference technique. The spectrograph has already achieved a stability of 3.7m/s in short-term time scale and in the near future we expect the stability to be at 1m/s once we install the spectrograph inside the vacuum chamber.

## 1. INTRODUCTION

The success of optical high-resolution spectroscopy using 1 to 2m class telescopes for detecting extra-solar planets has been substantial over the recent years and has revived the importance and usefulness of small size telescopes in the age of large and extremely large telescopes (Santos et al 2000; Queloz et al. 2000; Raskin et al. 2004, 2008). The window for small telescopes relies in the fact a) a very large number of stars up to 12$^{th}$ magnitude are yet to be monitored with high resolution spectroscopy for exoplanets studies, stellar seismology and pulsations, and other related sciences like abundance measurements, stellar rotational velocities, and b) a very large number of nights that are required for such science endeavors can only be available on 1 to 2m class telescopes.

Today some ~400 exo-planets have been discovered however, a number of basic/fundamental science questions remain unanswered. Small 1 to 2m aperture telescopes equipped with highly stable fiber fed spectrographs are required for significant contributions since discovery and characterization of exoplanet systems requires high precision radial velocity time-series spanning from weeks to decades. The full characterization of many exoplanets may be achieved using such long time based observations. Fundamental questions such as a) the range of planetary system architectures, b) statistics of planets in the habitable zone and c) formation of planets and planetary systems can be addressed once we began to observe host stars over very long period of time line.

[1] abhijit@prl.res.in; phone 91-79-2631-4607; http://www.prl.res.in/~abhijit

We at PRL have started our own exoplanet program using the technique of precision Doppler radial-velocity measurements. For this we have designed and built an optical fiber fed high-resolution echelle spectrograph, which is attached to our 1.2m Telescope at Mt. Abu, India. The observatory typically enjoys about 200 observable nights in a year out of which about 150 are photometric in nature. For our exoplanet program we have about 80 to 100 nights in a year.

Here we report the commissioning of the spectrograph in a temperature control environment, the first light data and the stability of the spectrograph, issues that we are facing and near future plan to solve the issues. PARAS stand for PRL Advanced Radial-velocity All-sky Search (see Chakraborty et al. 2008). The Echelle spectrograph was built in late 2009 and early 2010 and the spectrograph was commissioned and saw first light in February 2010.

# 2. SPECTROGRAPH DESCRIPTION

## 2.1 Optical Layout of the spectrograph and design considerations

The spectrograph is designed with an aim of extreme stability for doing precision radial velocity measurements down to a few meter/sec (<3m/s) with a basic spectrograph resolution of R~60000 between 3700A to 8600A. Its an optical fiber fed R3.73 (blaze angle of 75°; made from Master MR160 from Richardson Gratings) Echelle white pupil spectrograph with a single prism as a cross disperser and three elements one singlet, one triplet and one doublet Camera lens system. Since the spectrograph is planned for a small telescope of 1.2m aperture care has been taken for minimal light loss in the optics train and the detector quantum efficiency. Details of the design considerations are given in Chakraborty et al. (2008), however, we have made some changes in the glasses of the Camera system for higher throughput in the blue wavelengths and also the logistics of practical availability of large amount of glass. The Glass PKZ52 Scott glass has been changed to S-FPL52Y Ohara glass. The spot sizes remained unchanged.

Table 1 below summarizes the basic characteristics and Figure 1 shows the optical layout. Figure 2 shows the photograph of the actual spectrograph aligned and installed inside the temperature controlled inner chamber. Also shown is the mounting of the Echelle and the Prisms. Inset in Figure 2 shows the Fiber Optics holder. The two fibers are fixed by New-Port miniature X-Y stage with locking screws, which then are coupled to the F/4 to F/13 optics.

Table 1. Basic parameters of the spectrograph

| Component | Characteristics |
|---|---|
| Pupil Diameter | 100mm diameter, F/13 for the off-axis Parabolas |
| Off-axis parabolic mirrors M1 & M2 | Zerodur Glass, Cut from 60cm diameter Parabola, M1 is 140.5mm x 260.5mm x (54mm, thickness), Off-axis 127.5mm +/-0.5mm, Radius 2595mm +/-0.6mm; M2 is 130.5mm x 260.3mm x (55.5mm thickness), Off-axis 158.5mm +/-0.5mm, Radius 2595mm +/-0.6mm;<br><br>Surface quality of M1 and M2 are: Surface Front Error (SFE) within the useful area ~11nm(M1), 18nm(M2) rms, Periodic errors are at ~2.5nm, Micro Surface Roughness at ~1.5nm; Reflectivity (R): Special Protective Silver Coating, 4000A to 9000A > 95.5% (98.5% max at 7500A), 3800A to 4000A > 92%; Manufactured and Coated by SESO |
| Flat Fold mirror (FM) | Zerodur Glass, 160.44mm x 16.44mm x (33mm, thickness), SFE in the useful area ~ 19nm, Periodic errors ~ 1.5nm, Micro Surface Roughness < 1nm; Reflectivity: Same as M1 & M2 mirrors; Manufactured & Coated by SESO |
| Echelle Grating | Dimension of 420.5mm x 220.5mm x 74mm; Useful ruled area 415mm x 215mm, Blaze angle 75°, used in Littrow condition, face down, with a small $\gamma$ angle of 0.45°;<br><br>Surface quality $\lambda/4$ P-V over the central region and almost 70% of the ruled area, this region is used by the 100mm diameter pupil, thus ensuring a higher spectral purity; Blaze peak up to 60% efficient for each order; Manufactured |
| | by New-Port Richardson Gratings |
| Prism | OHARA, PBM8Y glass, apex angle 65.6°, physical size of entrance and exit faces are about 226mm x 170mm; Prism Height 190.8mm; Transmitted Wave-front Error ~ 72nm rms; Surface roughness within the useful area ~1.5nm; Throughput efficiency measured at ~81% between 3800A and 8500A including both surfaces; Cut, polished and AR-coated by SESO |
| Camera Lens System | F/5, Focal Length ~528.5mm at 6330A, EE80 (diameter) on-axis and at +/- 3.1° field ~ 9microns, 96% of encircled energy with in 25microns; Average transmission thru the Lens system (total six surfaces) ~ 95% between 4000A and 8600A; Manufactured, assembled and Zygo Interferometer test is done by SESO |
| CCD Detector | E2V, 4096x4096, 15micron square pixel, CCD231-84-1-D81, Deep Depletion, Astro-Broad-Band coating, measured QE ~ 57% at 3500A, 92.4% at 4000A, 96.5% at 5000A, 88% at 6500A, 52% at 9000A; gain ~1.7, slow readout read noise ~ 3.5electrons, Dark noise ~ 0.02electrons/pixel/hour at -120°C temperature;<br><br>CCD Dewar and cryogenics is assembled by Infrared Laboratories. The IR-Lab Dewar operates at -120°C using low vibration (<150nm vibration amplitude on the Detector) Helium closed cycle Cryo-cooler. The Low vibration Cryo-Cooler DMX-20 head is from Advanced Research Systems, Inc., and the CCD controller is from Astronomical Research Cameras, Inc., Data acquisition and communication between the Controller and the PC is done thru two-way fiber optics communication and the PC runs Voodoo software under Red-Hat Linux. The Science Grade CCD was installed inside the Dewar by the author (AC) at PRL. |
| Fiber Optics | Multi-mode, FBP type, core 50microns, including clad ~85microns, 20m in length, very good transmission over 20m length, only up to 15% light loss due to the length of the Fiber; The Star fiber and the Calibration fiber form a bundle on the spectrograph side such that these two fiber cores are separated by 180microns +/- 3microns which corresponds to about 17 pixels separation of the spectra on the Detector<br><br>The 1.2m telescope F/13 beam is converted into F/4.5 using a Focal reducer and the 50micron star fiber tip directly inserted in the F/4.5 telescope focal plane, the fiber see about 2-arsec on the sky; however on the spectrograph side F/4 beam coming out of the fiber tips are first collimated by a F/4 doublet and reimaged on to the slit position using a F/13 Doublet; both the doublets are custom made achromats defining spot sizes within 20microns diameter; the Calibration fiber has the same configuration of Lenses at the other end |
| Projected 50micron fiber diameter at the slit position and resolution R | ~170microns corresponding to Resolution (R) ~63000; there is no slit at the slit position. The projected fiber image defines the resolution, thus avoiding slit losses and errors in precision radial velocity |
| Tip/Tilt unit | For Image stabilization at the fiber tip, and for correcting telescope tracking errors, in transmission, 12mm thick and 40mm diameter, AR coated BK7 glass plate, +/-150microns max displacement corresponding to +/-3° tilt; Installed in the F/13 beam of the telescope; The tip/tilt unit usually operates at 1 to 2Hz frequency and sensitive up to $10^{th}$ mag star, but can also operate on very bright stars up to speed of 10Hz. In the future we plan to move it in the F/4.5 beam after the focal reducer (see section 4.1) |

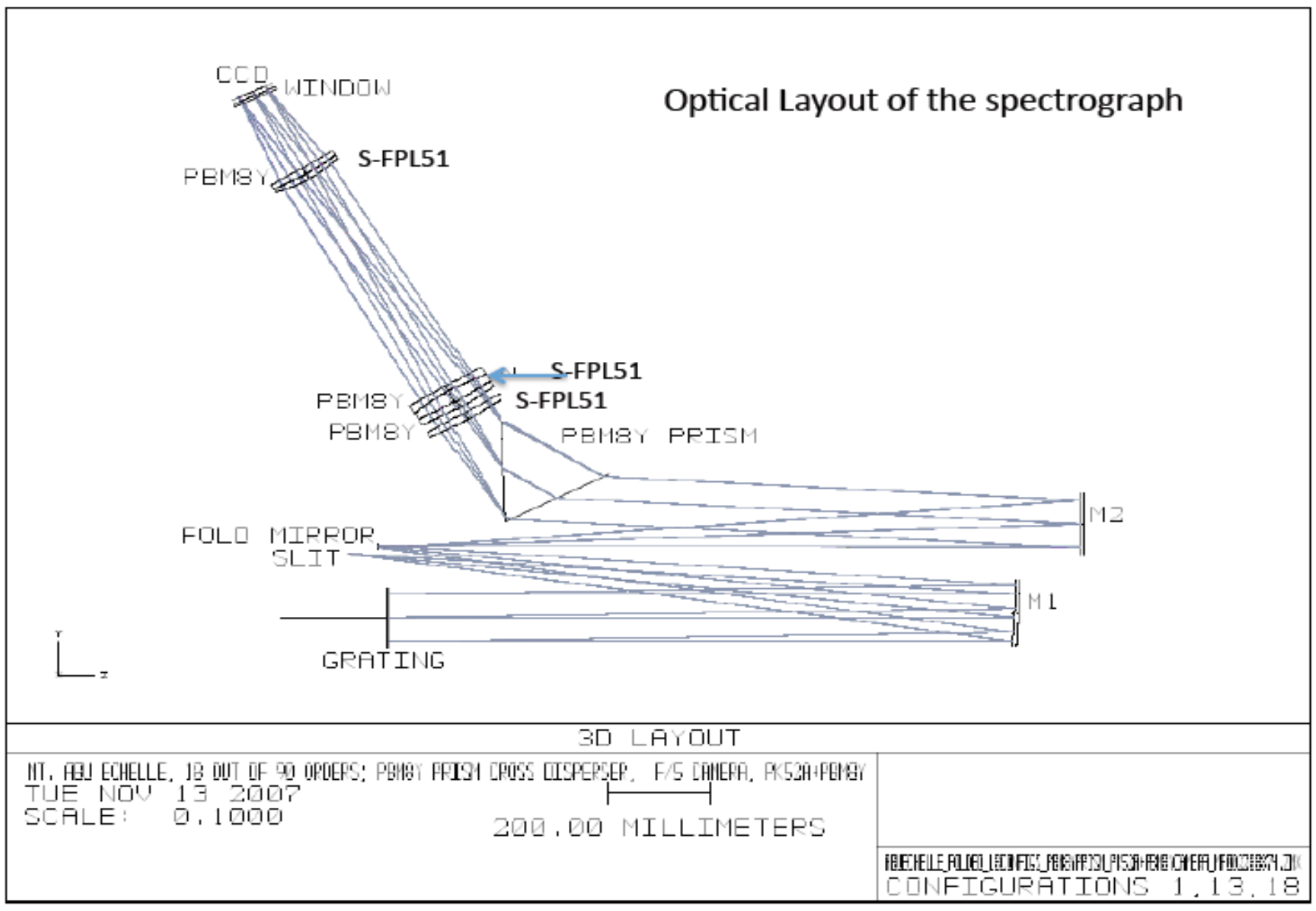

Figure 1. The Optical Layout of the Spectrograph

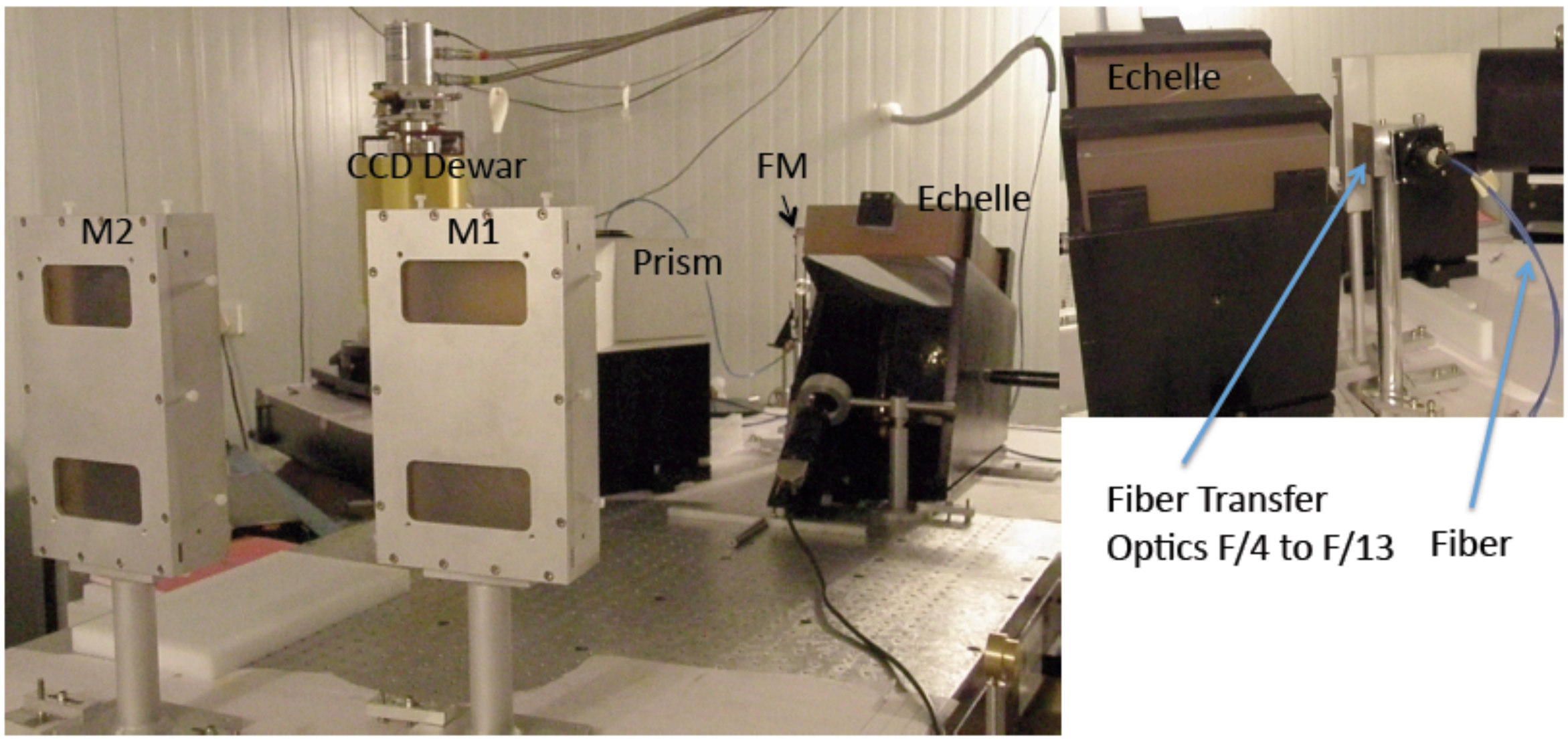

Figure 2. The PARAS spectrograph optics aligned on the optical bench in a temperature control environment, where M1 & M2 are Off-axis Parabolic Mirrors and FM is the Fold Mirror (Flat). Inset: 2 Fibers (star + calibration) coupling into the spectrograph

Thus, the spectrograph is estimated to be ~30% efficient from the slit position to the Detector Q.E. at blaze peak wavelengths and if the fiber optics, fiber transmission, telescope reflectivity and the Mt. Abu sky is included then under photometric conditions the overall efficiency of the spectrograph is estimated to be around 10%. This is off course subject to telescope tracking errors and flexures (see section 4.1).

## 2.2 Temperature control

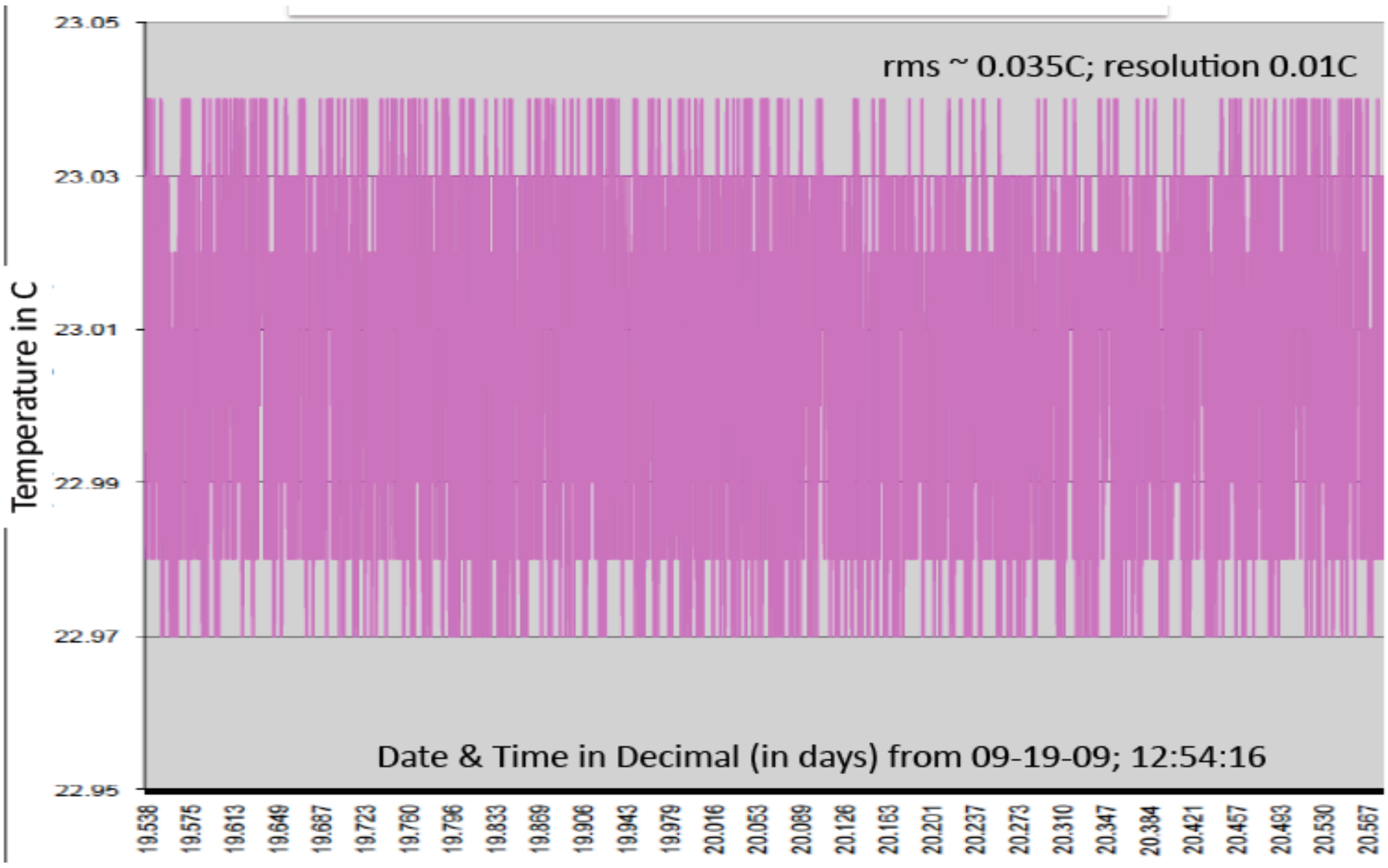

Figure 3. A sample of temperature variations in the inner Chamber. Here the Graph shows the variation over 24 hours of time.

The spectrograph is installed in a closed concentric volume space consisting of the outer and the inner chambers. The optical bench is kept on vibration isolation RCC pillared (isolated from the observatory building by a few cm gap from all sides) polished Granite top table. The inner chamber is built around this special table. The walls of the chambers are made up of highly insulated 60mm thick puff material (materials similar to those used in standard commercially available refrigeration systems). The outer chamber temperature control is achieved at 20°C +/-0.8°C using standard Heating and Cooling mechanisms, which are commercially available. It is PID controlled and the temperature can be set at any value between 18°C and 24°C. We have tested it at 21°C and at 24°C with a precision of +/-0.8°C.

The inner Chamber has an inbuilt temperature control system. We control the temperature by increasing it about 2°C above the outer chamber temperature. Electronic Thermal controlling is used to control the heaters and record the data. We have used Pt100 temperature sensor, which is well known for its high stability, ~ 0.005% or better than that. Pt100 along with associated pre-amplifier convert the temperature into equivalent analogue values. The ADC (ADC7135) converts this analog information from the sensor to digital information with a resolution of 0.01°C. In the near future we plan to increase this resolution to 0.001°C. ADC reference voltage is highly stable. We have used REF200 from Burr-Brown. The sampling frequency is around 100khz. ADC output is compatible and connected with one of the port of microcontroller. We have used Atmel 89c51/52 microcontroller to digitize the temperature information and sending appropriate data to control the heater. A preset temperature value is required to enter to the microcontroller. The microcontroller compares this preset value with the current temperature. If there is a difference then it passes the appropriate digital value to heater circuitry and to heater power driver. The program is written in assembly language to display the temperature on LCD, and also simultaneously it also sends the temperature data Serially to PC through Comport (RS232). Data Acquisition Software is written in VISUAL BASIC to acquire the data and storing it in hard disk. Figure 3 demonstrates the temperature control of the inner chamber at 23°C when the outside chamber was controlled at 21°C.

## 3. FIRST LIGHT STELLAR SPECTRA FROM PARAS UNDER TEMPERATURE CONTROL ENVIRONMENT

Before installing the PARAS spectrograph inside the vacuum chamber we decided to align the spectrograph on an Optical bench which is identical to the shape and size of the vacuum chamber base plate in the temperature controlled environment. The basic idea was to build and test the characteristics of the spectrograph and see if it met most of the design parameters before pushing it inside the vacuum chamber. We also wanted to test the stability of the spectrograph under the temperature control environment alone using the simultaneous Th-Ar calibration technique (Barnes et al 1996; Pepe et al. 2000).

From the telescope Cassegrain unit two fibers come to the spectrograph, one the Star Fiber and the other the Calibration Fiber as mentioned in Table1. The Cassegrain unit also contains the calibration unit, which has the Th-Ar spectral lamp, Flat Lamp, ND filters for brightness control and F/13 beam simulator for the calibration light sources. This F/13 beam can be fed to both the fibers and the star fiber can receive the F/13 Calibration beam instead of the F/13 telescope beam when desired. The design is such that the F/13 calibration beam then passes thru the F/4.5 focal reducer and forms a spot image on the Star fiber. The Calibration Fiber always receives the F/13 calibration beam thru an F/13-F/4 reimaging system but can also be blocked off using a shutter when required. Figure 4 shows one such calibration image when both the fibers are illuminated with Th-Ar light. The spectrum shows that we have achieved resolution of 63000. The space between the two fibers spectra of same order remains constant across the echelle orders and it is about 17pixels. Thus, we can take simultaneous Th-Ar spectra on both the fibers, or star spectra on star fiber and Th-Ar spectra on the calibration fiber from 3700A to about 6700A, beyond that the different echelle orders comes too close for both fibers simultaneous imaging. This is acceptable because the precision radial velocity observations are primarily done between 4000A and 6800A. For non-radial velocity sciences where the longer red wavelength region are important one can observe single fiber (star-fiber only) exposures and can achieve wavelength calibration by taking calibration spectra before and after the science observations.

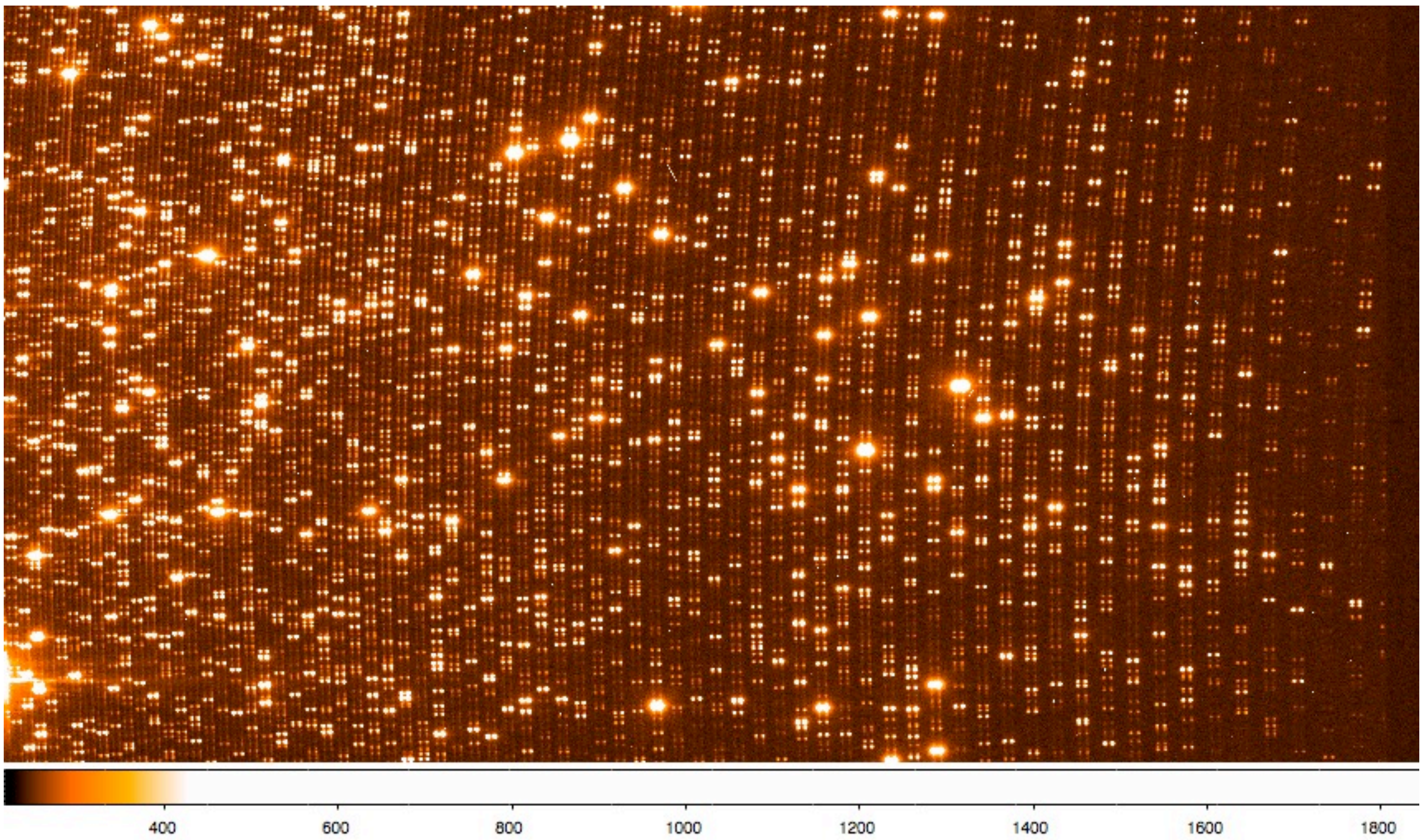

Figure 4. Zoomed Part of the raw Image showing Echelle orders of the Th-Ar spectra on both the Fibers (Star Fiber and Calibration Fiber) between 3700A and 6800A; Spectra shows that R~63000 has been achieved; The full image image goes from 3700A to 8600A.

Th-Ar spectra on both the fibers if taken at regular intervals determine the stability of the spectrograph (see Figure 5). Over the night we typically achieve about 4m/s of stability. Figure 5 shows stability of 3.7m/s over the night. The prime issue that we faced in the stability of the spectrograph is the flexing of the optical bench. Even though the optical bench is installed on a vibration isolated Granite table top, there are thermal insulators between the firm Granite base and the Optical bench, which themselves were differentially flexing during the night of observations causing very large absolute drifts in hundreds of meters/sec. Tracing the absolute drifts and recovering them is tricky business and hence we achieved about 3 to 4m/s spectrograph stability. Since we have identified the problem during the vacuum chamber installation process (see next section; which is going on at the time of writing this manuscript), we now plan to install Trisolators between the vacuum chamber and the firm Granite Vibration isolated platform. The thermal isolation between the Granite and the chamber will be done thru special Trisolators, which are also primarily vibration isolators. We expect to reach 1m/s stability after this particular operation.

Figures 6 and 7 show small sample spectra of stars 47 Uma (G1V, $m_v$ ~ 5.1mag), and tau Boo (F9 IV, $m_v$ ~ 4.2mag) obtained on 1st April 2010. Both these planets are known to host exoplanets. The stellar rotational velocity of the 47Uma is ~2.8km/s, while that of tau Boo is ~14km/s, which is very clearly evident from the spectra of the stars. The various FeI, features TiI are seen in the spectra and can be identified. Thus the basic spectrograph is working for non-radial velocity sciences like for stellar abundances measurements etc. Meanwhile, we are working on the cross-correlation data-pipe line issues for obtaining precision radial velocity on stars including developing G2 mask for the PARAS spectrograph.

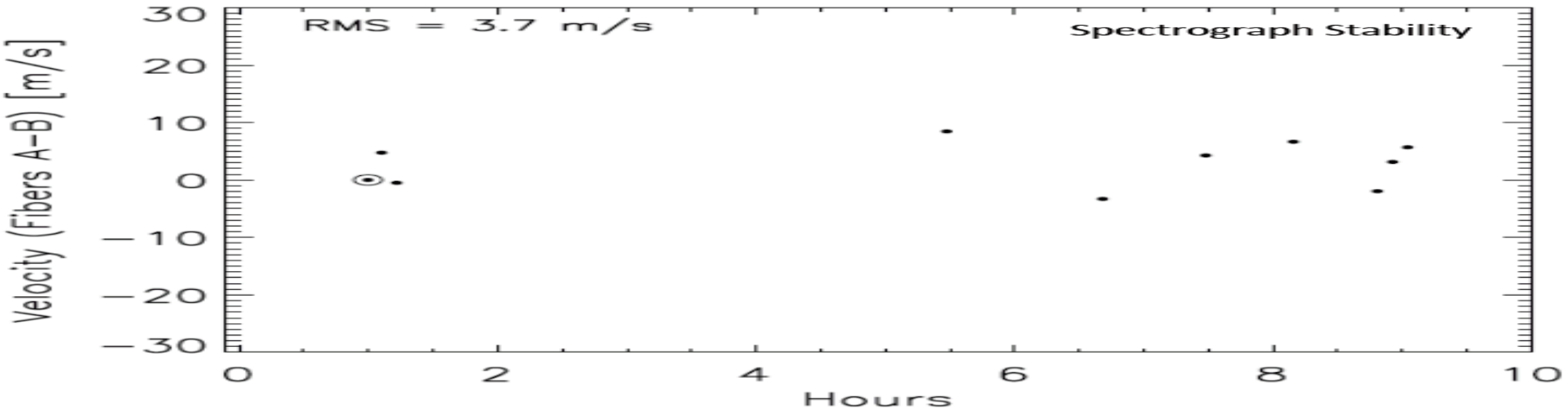

Figure 5. Spectrograph stability of 3.7m/s over the night, it is obtained by taking the absolute difference of radial velocity instrument drift measurements between the two fibers A (star fiber) and B (Calibration Fiber). In this case both the fibers are exposed to the Th-Ar Spectral Lamp light like in Figure 4.

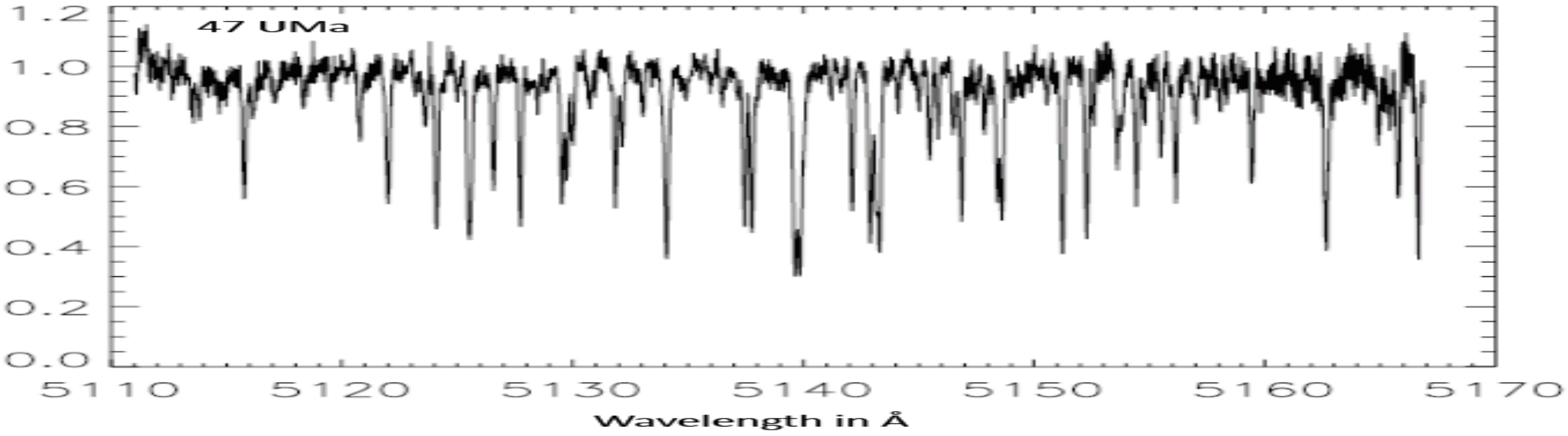

Figure 6. A small sample Spectra of 47Uma, which was observed on 1st April 2010. The stellar rotational velocity is 2.8 km/s

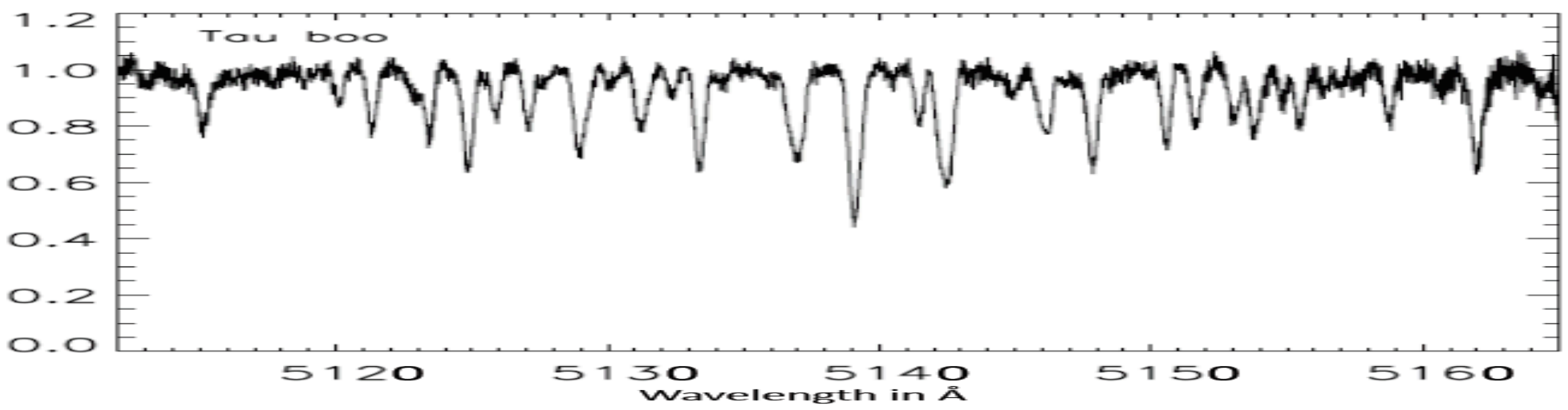

Figure 7. A small sample Spectra of tau Boo, which was observed on 1st April 2010. The stellar rotational velocity is 14km/s

# 4. FURTHER DEVELOPMENTS IN THE NEAR FUTURE

### 4.1 A new F/4.5 Focal Reducer with the Tip/tilt unit in the F/4.5 beam for efficient star tracking on the Fiber tip

One of the primary issues we are encountering while coupling the fiber optics with the telescope is that of the telescope flexure between the Primary mirror and the secondary mirror and also the telescope tracking error. Since the telescope runs on an old technology, it has tracking accuracies of only 2arcsecs and suffers from hysteresis in RA and DEC drives and also huge flexures and hence alignment issues between the primary and the secondary. Thus, due to flexure alone, the star image moves in the focal plane by several hundreds of microns as it tracks the star from East to the West. The present Tip/Tilt unit which works in transmission and has maximum displacement of +/-150 microns and on the telescope F/13 beam it corresponds to only +/-2arcsecs, which is same as that of the telescope minimum movement. As a result we found that star oscillates around the fiber tip with amplitude of 2.5arcsecs and causing significant light loss of factor 10.

Therefore, we now plan to insert the Tip/Tilt unit in the F/4.5 beam to give us a maximum throw of +/-6arcsecs and will operate it at 1-2Hz frequencies. This will take care of both the issues a) tracking errors with hysteresis and the telescope flexure issues, and star should be stable at sub-arcsec level on the fiber tip. However, the present F/4.5 focal-reducer does not have enough physical spaces between the last lens surface of the focal reducer and the fiber tip for the Tip/tilt unit with ~15% Pellicle beam splitter to be inserted in. We have designed a new Focal reducer which has about 150mm physical space between it and the fiber tip for the installation of the Tip/Tilt unit. With this change we hope to achieve S/N ~50 on 8th mag star in less than 10mins of exposure and in about 40mins of exposure time we should be able to reach S/N >70 on a 10th mag star. The Tip/tilt unit is Starlight-Xpress make.

### 4.2 The Vacuum Chamber

In the near future we plan to move the spectrograph inside a vacuum chamber for pressure control and ultra-high stability (the installation process has started at the time of writing this manuscript). The Vacuum Chamber is made of Stainless Steel 304L material and has large rectangular port openings for easy access to optical components placed inside the chamber. The ports are covered using flanges made of Aluminum. Both the materials are vacuum compatible and can hold vacuum with vacuum pumps isolated, due to low out-gassing properties. The shape of the vacuum chamber was optimally chosen to enclose the entire optical path from the slit position to the CCD Dewar. Thus one axis of the chamber is about 2 m in length while the other is about 1.6 m. The height of the vacuum chamber is about 0.7 m and is designed to enclose partly the CCD dewar, leaving the Helium line connections for cryo-cooling outside the vacuum chamber. This arrangement is done through flexible stainless steel bellows. The base plate of the vacuum chamber acts as the optical bench and has a surface flatness of 20 microns. The optical axis is 300 mm above the optical bench.

The Starlight and the Calibration light will get into the Vacuum chamber thru two separate optical window ports. The optical windows are AR coated. The fibers in air (outside the chamber) will couple with the fibers in side the chamber (in vacuum) thru optical fiber scrambler described in the next subsection. The electrical connections are taken into the vacuum chamber through vacuum compatible multi-pin feed-throughs.

The base plate and the port flanges are adequately strengthened by external ribs, and the chamber walls are strengthened by internal ribs to give minimum deformation under vacuum. ANSYS CAE (Computer Aided Engineering) Software Program was used in conjunction with 3D CAD (Computer Aided Design) solid geometry to simulate the behavior of mechanical bodies under thermal and structural loading conditions. ANSYS automated FEA (Finite Element Analysis) technology was used to generate the results. The analysis was done for static structural deformation and stress at three uniform ambient temperature 0°C, 20°C, and 40°C for an external pressure of 1 atmosphere. The maximum deformation was ensured to be within 50 microns and maximum stress was ensured to be well within the minimum yield strength. Aditya High Vacuum manufactured the Chamber.

The evacuation of the Chamber is done using a Dry Vacuum Pump (Alcatel make, Multi-roots type) to a vacuum less than $10^{-1}$ to $10^{-2}$ mbar. The vacuum is measured using an absolute pressure measurement gauge using capacitance manometer principle. Liquid Nitrogen cooled Activated Alumina Trap improves the vacuum to the range of $10^{-2}$ to $10^{-3}$ mbar and also allows to hold the vacuum in the chamber at a steady value even after isolation of the Displacement pump. The fabrication of the vacuum chamber is completed and tested for holding the vacuum at $10^{-2}$ mbar for over 36 hours. We have also tested the actual deformations and verified them against the theoretical ANSYS results, and they are as per expectation within 30microns and are well within the limits. We have also tested that as long as the holds the vacuum of $10^{-2}$ mbar the deformations remains constant for the stability of the spectrograph.

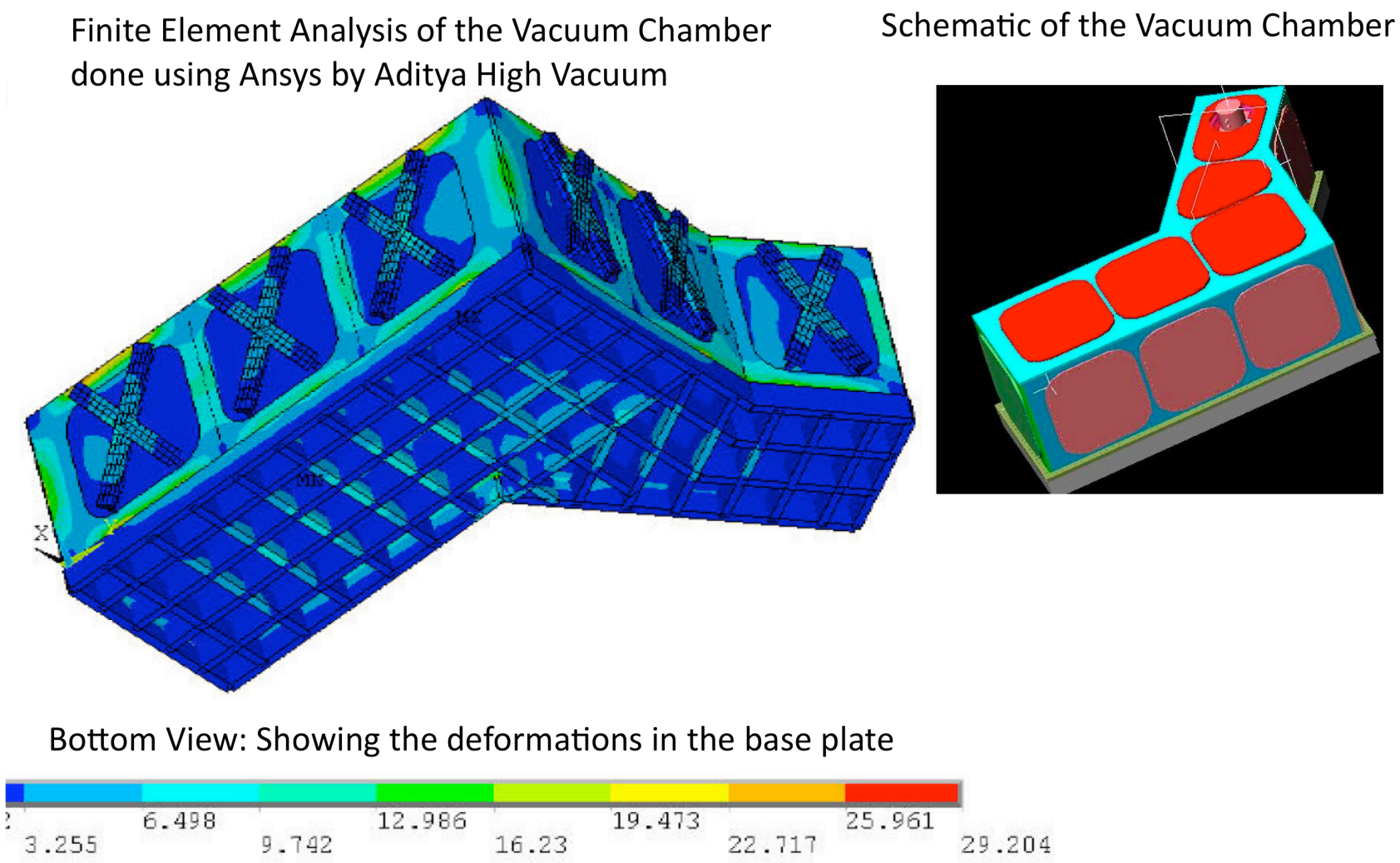

Figure 8. Deformations in microns with 1 atm. pressure difference at 25°C. The deformation remains stable as long as the pressure remains stable. Typical pressure stability has been tested at $10^{-2}$ mbar.

### 4.3 Optical Scrambler for high precision radial velocity down to 1m/s

One of the major applications of fiber-coupled spectrographs is high precision stellar radial velocity observation for exoplanet searches. Non-uniform or varying illumination at the slit of the spectrograph decreases radial velocity precision and so it is desirable for obtaining precision <10m/s that the star image or the fiber illumination at the slit position should remain constant. Typically, telescope tracking errors and/or atmospheric seeing variations can introduce such variation in the illumination. Fibers fortunately can scramble the structure of the input star image and can produce a uniform structure image of the output beam. However, the properties of cylindrical Fibers are such that they can scramble the input image effectively more azimuthally. That is the output beam is symmetrical about the axis of the fiber, and the degree of the input variation will appear as variations in the radial direction in the output beam (Hunt Ramsey 1992). For instance if light is entering the fiber at an angle then the output beam will show an azimuthally symmetric ring like structure in the output beam (see figure 9 right (a)).

The issue of radial non-uniformity is solved by the introduction of a double Scrambler first introduced by Hunter & Ramsey (1992). Later on the principle has been applied to spectrographs like *HARPS* (Mayor et al. 2003, Pepe et al. 2002) & *Sophie* (Perruchot et al. 2008) for precision radial velocity measurements.

We have designed our own Double scrambler (see Figure 9, left), which consists of two doublets. Each doublet is of 2.8mm diameter and is made up of N-BK10 and FPL53 Glasses. As shown in Figure 9(left), the doublet on the left side is sitting on the Fiber tip, which is coming from the telescope, and the doublet on the right side is sitting on the fiber tip, which is going to the spectrograph. The doublets convert all rays coming out from the first fiber at different angles to get at different positions on the second fiber, and conversely all rays coming out from different positions from the first fiber enter the second fiber at different angles. Thus azimuthally symmetric light cone coming out of the first fiber enters radially in to the second fiber. We have tested this concept with the doublets in the Lab as shown in Figure 9 (right). Figure 9(right)a shows the output beam from the Fiber from the left where the ring like structure is due to light entering the fiber at an angle (~simulation for telescope tracking error). Figure 9(right)b shows the output beam from the second fiber which is uniformly symmetric on both the axis radial and azimuthal.

We plan to install the Double scrambler optics at the Vacuum Dewar Interface along with the installation of the Vacuum Chamber in the near future.

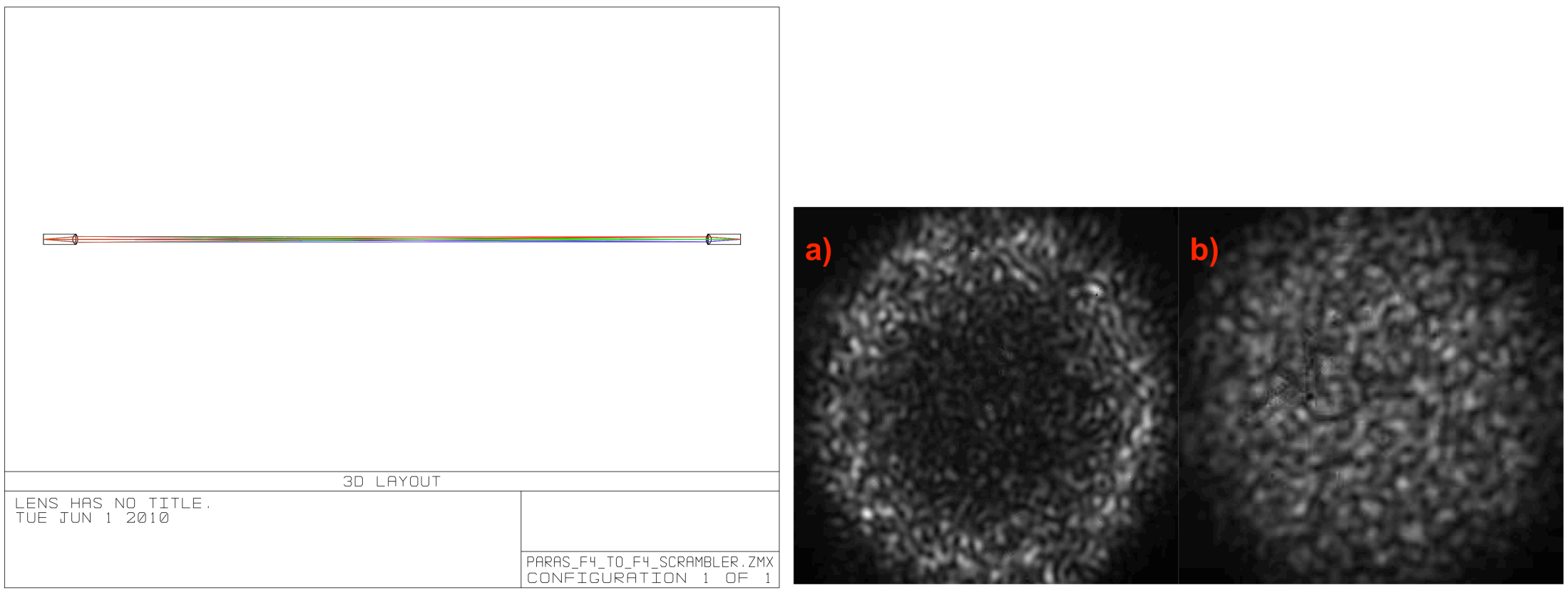

Figure 9. Scrambler Optics ray diagram. BMV Opticals based on our design manufactured the Scrambler Doublets. Right: the effect of scrambling, even though the input beam appears to be in circular ring like structure, the output beam will always appear uniform.

## Acknowledgements

The authors would like to acknowledge J.N. Goswami, the director of Physical Research Laboratory, and Dept. of Space, Govt. of India for funds and grants for the support of the PARAS project. AC would like to thank Francesco Pepe from Geneva Observatory and Larry Ramsey from Penn State University for many valuable inputs and their active interests in the project. AC would also like to thank the staff of the Observatory and PRL administration for their active co-operation during various installation processes. AC would also like to acknowledge various faculty members at PRL, particularly B.G. Anandarao & N.M. Ashok for helping in various logistics since the project began in 2007. SM would like to thank American Astronomical Society (AAS) for the travel grant to travel from US to India and back in January 2010. Finally, AC would like to acknowledge D. Subrahmanyam, J. Desai, A. Hait & R.V. Nair from Space Application Centre, ISRO, and Venkat Ramani from Aditya High Vacuum for their valuable inputs on the basic design of the Vacuum Chamber.